\documentclass{svproc}
\usepackage{graphicx}%
\usepackage[english]{babel}
\usepackage{multirow}%
\usepackage{amsmath,amssymb,amsfonts}%
\usepackage{mathrsfs}%
\usepackage[title]{appendix}%
\usepackage{xcolor}%
\usepackage{textcomp}%
\usepackage{manyfoot}%
\usepackage{booktabs}%
\usepackage{algorithm}%
\usepackage{algorithmicx}%
\usepackage{algpseudocode}%
\usepackage{listings}%
\usepackage{subcaption}%
\usepackage{hyperref}
\usepackage[table]{xcolor}%
\usepackage{url}

\usepackage{cite}
\usepackage{paralist}
\usepackage{xspace}
\newcommand{\eg}{e.g.,\xspace}
\newcommand{\ie}{i.e.,\xspace}

\newcommand{\eat}[1]{}

\begin{document}
\mainmatter              
\title{Statistical Feature Augmentation for\\ Anomaly Detection in Dynamic Graphs}
\titlerunning{Statistical Feature Augmentation for Anomaly Detection in Dynamic Graphs}  
%
\author{Philipp Schlinge\inst{1} \and Jean-Luc Schnipper\inst{1}
	\and Martin Atzmueller\inst{1,2}}
\authorrunning{Philipp Schlinge et al.} 

\institute{Osnabrück University, 49074 Osnabrueck,Germany,\\
	\email{\{philipp.schlinge,jschnipper\}@uos.de},
	\and
	German Research Centre for Artificial Intelligence (DFKI),\\
	49084 Osnabrück, Germany\\
    \email{martin.atzmueller@uos.de}}

\maketitle              

\begin{abstract}
	Dynamic networks are being applied in many domains, from social media to logistics systems, each with their own set of special characteristics. A
	model employed on this type of data must capture the duality between temporal/structural and feature-based information. Yet
	state-of-the-art deep learning models often struggle to learn especially short-term behavioral interaction signals, such
	as sender intensity or interaction inertia, directly from raw event streams. To address this gap,
	we propose a statistical feature augmentation method that explicitly encodes behavioral interaction
	statistics into the input feature space. We
	evaluate our proposed method on an anomaly detection task across three real-world datasets (Reddit, Wikipedia, MOOC)
	and seven models spanning both continuous-time and discrete-time architectures. As a baseline, we apply the same models trained on the original embeddings. Our results show, that augmentation consistently improves detection performance. Beyond 	performance, the enriched input enables fine-grained post-hoc analysis of behavioral importance, since each statistic occupies a dedicated input dimension. In particular, this
	work showcases a promising approach for merging classical network analysis with deep learning.\\
    Code is available at: \href{https://github.com/uos-sis/base-stat-aug-ad}{https://github.com/uos-sis/base-stat-aug-ad}
\end{abstract}
\section{Introduction}\label{sec1}
Deep learning models for dynamic graphs can be categorized into discrete-time models leveraging
static graph learning frameworks like GraphSAGE~\cite{graphsage} and continuous-time models that
focus on sequence learning designs, for example DyGFormer~\cite{dygformer}. The former employs
discrete snapshots aggregated over a fixed time interval as the data format, while the latter
directly operates on the continuous event stream. These representations impose different structural
and temporal constraints on learning, motivating distinct architectural choices such as graph
convolution, memory modules, or temporal neighborhood sampling. Each of these choices targets a
specific subcategory of interaction patterns, capturing the behavior within those, \eg when considering a directed dynamic graph and the respective interactions between senders and receivers of events.

Despite this diversity, both model families struggle to reliably learn short-term behavioral characteristics such as sender intensity or interaction inertia.
Recent proposals to encode
comparable interaction patterns focus on complex motif learning strategies deeply embedded in the
architecture, \eg through random walks or other discovery mechanisms~\cite{neurtws}. In network analysis,
heavily structured models like relational event models (REMs)~\cite{rem} explicitly exploit such
statistics, but are restricted to transitive feature combinations and relatively simple models. Recent extensions, \eg STREAM~\cite{stream} and DREAM~\cite{dream} aim to integrate deep learning into the REM framework, but remain restricted by the model structure.

In this paper, we address this gap between deep learning and explicit statistical modeling by augmenting the input feature space with temporally-decayed behavioral statistics from classical social network analysis. This yields short-term information directly available to arbitrary deep learning architectures and creates dedicated input dimensions that can be propagated in post-hoc analysis.
To evaluate our proposed approach, we choose anomaly detection, motivated by the assumption that short-term interaction characteristics carry high discriminative information
in this setting. We cover a broad range of models spanning continuous-time and discrete-time
architectures.  With individual input dimensions for statistics, our setting facilitates the use
of general post-hoc explainability methods such as SHAP~\cite{shap} to quantify the influence of
individual behavior characteristics on model decisions. This also addresses the often under-explored
task of explaining the characteristics of predicted anomalies.

As an example, let us consider a discussion forum in the context of a Massive Open Online Course (MOOC): If a student $u$ typically posts one question per day to a forum but if at time $t$ she suddenly sends 20 messages to student $v$ within a span of two minutes up to $t'$, then this can indicate anomalous interaction behavior. 
Here, discrete-time models might aggregate these events into a single snapshot, in consequence losing the burstiness of the interaction. 
Continuous-time models, however, are able to capture the specific timestamps but rely on learned embeddings to infer that this burst is anomalous, which often requires a large set of training data to recognize such short-term patterns. 
In our example, we can observe deviations from the user's historical interaction intensity, with a sudden change in rate, as well as the inertia considering respective edges $(u,v)$ during the time interval $[t,t']$.
Without augmenting the input by explicit features, the model must thus implicitly learn these temporal statistics. In addition, during post-hoc analysis, a method like SHAP can directly attribute the anomaly score to the high value of suitable features, for example, intensity, and can thus provide immediate interpretability in contrast to only considering raw embeddings.

Our contributions are summarized as follows:
\begin{inparaenum}[(1)]
\item We propose and formalize statistical feature augmentation approach for anomaly detection in dynamic graphs.
\item We show that our proposed augmentation approach increases anomaly detection performance, often by a considerable margin.
\item We demonstrate that moving classic interaction statistics into the input feature space, substantially improves explainability of the studied anomaly-detection models.
\end{inparaenum}

\section{Related Work}\label{sec2}

Below, we discuss related work on dynamic interaction networks, statistical network models and motifs as well as anomaly detection in this context. Furthermore, we discuss explainability, in particular focusing on post-hoc explanations.

\subsection{Dynamic Interaction Networks}
A network is often modeled as a graph consisting of a set of \emph{nodes} and a set of \emph{edges} connecting the nodes.
For example, social networks represent people or groups of people (nodes) and relationships between them (edges).
Social interaction networks~\cite{WF:94} model \emph{interaction} relations between actors represented by \emph{people}. We can then empirically study such networks to gain an understanding of their underlying structures.
Furthermore, we can also consider other relationships which are mediated via specific resources. According to this, the principle of object-centric sociality~\cite{KC:97} then induces  connections between actors, based on implicit user interactions on specific joint resources.
The analysis of social interaction networks is discussed in~\cite{Atzmueller:14:CoRR}, specifically focusing on dynamic networks captured during certain events such as conferences, \ie as human behavioral (offline) networks.
The analysis of dynamic interaction networks -- as a form of feature-rich networks~\cite{interdonato2019feature}, also potentially including attributes -- represents a considerable evolution beyond static interaction models by explicitly integrating the timing, duration, and sequence of connections. Static networks aggregate interactions into a single topology. This can obscure the causal order necessary for understanding complex interaction processes. Dynamic frameworks preserve the time-resolved structure of edges, allowing for the identification of transient structures such as motifs and communities which could otherwise by masked~\cite{holme2012temporal}. Hence, this is the scope of this paper, focusing on dynamics and the respective time-dependent characteristics and features.

\subsection{Statistical Network Models and Motifs}
Statistical analysis of dynamic networks~\cite{snijders2001statistical,carley2007toward,goldenberg2010survey}, has a long history, with classical methods such as
Relational Event Models (REMs)~\cite{rem} that model the characteristics of a network through
point processes. Recent extensions increase the expressiveness from such REMs by introducing non-linear effects modeled through splines (STREAM~\cite{stream}) or neural networks (DREAM~\cite{dream}), but remain tied to the
statistical modeling paradigm. Complementary to these generative approaches, motif discovery or
motif learning such as DyG2Vec~\cite{Alomrani2022DyG2VecRL} are established methods for capturing characteristic
interaction patterns. Techniques in this line also include motif retrieval through random
walks~\cite{caw,neurtws} or discovery through statistical hypothesis testing~\cite{sitgnn}. This paper proposes an approach augmenting the input features with characteristic behavioral features, adressing the gap between deep learning and explicit statistical modeling. 

\subsection{Anomaly Detection on Dynamic Graphs}
From an abstract point of view, an anomaly is defined as a pattern that does not conform to some notion of the expected, normal behavior. Therefore, a straightforward general anomaly detection approach defines a region covering the expected behavior and declares any observation in the data that does not belong to this region as an anomaly.
For static networks, respective methods are discussed in~\cite{Savage:14,Akoglu:15}.
In contrast, anomaly detection in dynamic networks is more challenging, and concerns a large research area, spanning both
continuous-time~\cite{tgat,tgn,dygformer} and discrete-time~\cite{graphsage,strgnn}, typically in supervised~\cite{aer} or
semi-supervised~\cite{sad} settings. Continuous-time models often rely on neighborhood
sampling with attention mechanisms~\cite{tgat,dygformer,sad} to capture the structural context,
while discrete-time models ingest snapshots and use for example graph convolutional layers~\cite{graphsage,strgnn}
from the static domain. Within this space, we position our work as a data-centric augmentation
strategy that is model-agnostic: it enriches the input features, not the architecture, and is
therefore applicable to any of the above models, thus providing a general approach.

\subsection{Post-hoc Explanations}
Explainability and  explanations~\cite{roth2008explanation,gunning2017explainable,AFKS:24} are important for enhancing interpretability and comprehension of complex models, tackled by methods in the area of Explainable AI (XAI). According to~\cite{barredo2019explainable}, XAI should produce details and/or reasons in order to ``make its functioning clear or easy to understand''~\cite{barredo2019explainable}. Here, specifically post-hoc approaches are relevant for many cases, since these are typically model agnostic.
For machine learning on graphs, and specifically for dynamic graphs, explainability remains both underexplored and loosely defined~\cite{AFKS:24}.
Approaches for motif-level explanations such as TempME~\cite{tempme} and
T-GNNExplainer~\cite{Xia2023ExplainingTG} are rather restrictive to dataset and model formats. A clear explanation of the found
motif is an additional challenge typically not covered in such methods.
Post-hoc feature attribution methods such as SHAP~\cite{shap} have seen wide adoption in static
deep learning models. These general-purpose explainers are applied to the input feature space,
which makes them useful for our augmentation as each statistic maps to a dedicated input
dimension. This results in an importance analysis over a clearly defined set of basic
motifs/statistics, which is a simple yet highly interpretable setting.

In summary, prior deep learning approaches encode statistical patterns only implicitly with complex architectures, and explainability remains an underdeveloped feature on most dynamic graph models.
By explicitly placing the statistics into the input space, our work offers a model-agnostic approach
that both improves detection and simplifies explanations. In this regard, we see our work as
complementary to the existing literature in the form of a sophisticated preprocessing approach.

\section{Statistical Feature Augmentation}\label{sec3}
We formalize the input to any downstream model as a sequence of interaction events
$\mathcal{E} = \{e_t\}_{t=0}^{T}$, where each event
$e_t = (s_t, r_t, t, \mathbf{x}_t)$ consists of a sender $s_t$, a receiver $r_t$, a timestamp
$t$, and edge features $\mathbf{x}_t \in \mathbb{R}^d$. Anomaly detection then amounts to
classifying each event as normal or anomalous. In the following, we describe how we augment
each event's feature vector to bring the statistical information into the input space.

The core hypothesis of this work is that short-term interaction characteristics -- such as how often a
sender interacts, how recent that interaction was, and whether the sender is repeating past
interaction patterns -- carry high discriminative information especially for anomaly detection.
Classical social network analysis defines these characteristics as statistics over the recent event
history. We encode them with an exponential time-decay function, so we have a continuous range where
recent events carry more weight.

The temporal weight assigned to an event occurring at time $u$ and
evaluated at time $t$ is defined as
\begin{equation}
	w(u,t)=\lambda e^{-\lambda(t-u)},
	\qquad
	\lambda=\frac{\ln(2)}{T_{1/2}},
	\qquad t\geq u.
\end{equation}
The parameter $T_{1/2}$ controls the half-life of the decay and thereby determines the relevance of
historical events for the current statistic. A smaller $T_{1/2}$ gives more weight to recent events;
a larger $T_{1/2}$ smoothes the statistic over a longer window. In our experiments we set $T_{1/2}$
to half the duration of the dataset, retaining substantial weight on a large part of the history, which keeps information of long range patterns intact.

To incorporate cyclic interactions into our statistical augmentation, we define the following notation. For an event $t$ associated with sender $s'$, let $p_m(t)$ denote the $m$-th preceding event involving $s'$, where
\begin{equation}
	p_0(t)=t,
	\qquad
	p_m(t)=\max\left\{u<p_{m-1}(t):s_u=s'\right\},
	\quad m\geq1.
\end{equation}
The set of events in which $s'$ changes its receiver is given by
\begin{equation}
	D_{s'}=
	\left\{
	e_t\in E:
	s_t=s',\;
	p_1(t)\text{ exists},\;
	r_t\neq r_{p_1(t)}
	\right\}.
\end{equation}
A cyclic interaction of sender $s'$ containing $k$ unique receivers is then
defined as
\begin{equation}
	C_k(t)=
	\mathbf{1}_{\{e_t\in D_{s'}\}}
	\mathbf{1}_{\{r_t=r_{p_{k}(t)}\}}
	\prod_{\ell=1}^{k-1}
	\mathbf{1}_{\{r_t\neq r_{p_\ell(t)}\}}
	\sum_{\ell=0}^{k}w(p_\ell(t),t).
\end{equation}

Table \ref{tab:dynamic_statistics} lists the considered set of statistics in our context.
Here, the intensity statistic captures the overall activity of a sender/receiver, while the inertia statistic quantifies the
recent activity on a specific sender-receiver edge. The interaction-cycle statistics capture whether
a sender changes its receiver in a way that completes a (directed) cycle of length $k$. These
statistics are computed per event based on the entire history up to time $t$.
Using the directed and undirected interaction case, we compute a set of 9 statistics per event.
\begin{table}[htbp]
	\centering
	\caption{The dynamic network statistics used for augmentation.}
	\label{tab:dynamic_statistics}
	\begin{tabular}{lc}
		\toprule
		Statistic                    & Equation \\
		\midrule
		Node Intensity                  & $I(i,t)=\sum_{j:(i,j)\in E}\sum_{e_u\in E_{ij}}w(u,t)$ \\[2ex]
		Edge Inertia               & $I(i,j,t)=\sum_{e_u\in E_{ij}}w(u,t)$ \\[2ex]
		Pair interaction cycle      & $C_2(t)$ \\
		Triad interaction cycle & $C_3(t)$ \\
		\bottomrule
	\end{tabular}
\end{table}

We consider two augmentation variants. The first, (+Stats), is the baseline of our augmentation and simply concatenates a normalized version of the statistics vector
$\mathbf{s}(t)$, to the original edge
features so that
\[
\mathbf{x}'_t = [\mathbf{x}_t ; \mathbf{s}(t)].
\]
This variant does not address asymmetries between the embedding feature space and the distributions
of individual statistics.
That is why, in our second variant, statistical feature encoding (+Stats +Encoded), we encode the
concatenated vector through a bottleneck autoencoder. The encoder maps
$\mathbf{x} \in \mathbb{R}^d$ through a sequence of hidden layers
into a latent vector $\mathbf{z} \in \mathbb{R}^k$; a decoder reconstructs
the input.

To make each latent dimension informative for anomaly detection, we optimize the
Kolmogorov-Smirnov (KS) separation between normal and anomalous distributions, together with a
reconstruction term and a variance penalty:
\begin{equation}
	\mathcal{L} = - \frac{1}{k} \sum_{d=1}^{k}
	\sup_{z} \left| F_n^{(d)}(z) - F_a^{(d)}(z) \right|
	+ \lambda_{var} \sum_{d=1}^{k} \left( \mathrm{std}(z_d) - 1 \right)^2
	+ \lambda_{rec}\, \mathrm{MSE}\!\left( \mathbf{x}, \hat{\mathbf{x}} \right),
	\label{eq:ks-autoencoder-loss}
\end{equation}
where $F_n^{(d)}$ and $F_a^{(d)}$ are the CDFs of the $d$-th latent dimension over normal and
anomalous samples, respectively. The KS statistic is computed with a differentiable,
temperature-smoothed CDF so it can be optimized via backpropagation.

The KS-Loss is suitable for our setting because each statistic encodes a distinct
behavioral dimension (intensity, inertia, cyclic activity), so we expect the latent dimensions
to be approximately disentangled. This assumption cannot be made for the compressed embedding
vector, which is why we use the reconstruction term to retain input information and the
variance penalty to prevent collapse or trivial scaling.
The computation of the actual statistics is efficient for large datasets: each statistic can be
updated in constant time per new event. This is enabled by the exponential decay weight
function: for a history of length $n$, the statistics over a node or edge are computed using
only $O(1)$ operations per step, yielding $O(n)$ total complexity for the entire stream.

\section{Experiments}\label{sec4}
We evaluate all models on the task of binary edge classification: given the historical event
stream up to time $t$, predict whether the current edge event $e_t$ is anomalous or normal.
We compare three feature configurations for each model: (1) the original edge features without
augmentation (baseline), (2) the original features concatenated with the statistics
(+Stats), and (3) the encoded latent representation from the KS-autoencoder (+Stats +Encoded).

We use three standard dynamic graph datasets (Reddit, Wikipedia, MOOC) commonly used in anomaly detection evaluation.
Table~\ref{tab:datasets} summarizes the properties of our dataset splits. All datasets provide ground truth anomaly labels for their events.
Each dataset is split chronologically into 70/15/15 for training, validation, and testing. All models
are chosen by their highest ROC-AUC performance on the validation split, and the final
performance is evaluated on the test split. The encoder model used to generate our (+Stats+Encoded) data variants is also trained on the training split and chosen based on the smallest loss on the validation split.

We evaluate seven anomaly detection models, spanning both architecture families from
Section~\ref{sec2}: continuous-time models (TGAT~\cite{tgat}, TGN~\cite{tgn}, DyGFormer~\cite{dygformer},
AER~\cite{aer}, SAD~\cite{sad}) and discrete-time models (GraphSAGE~\cite{graphsage}, StrGNN~\cite{strgnn}).
%
For the discrete-time models, we convert the continues edge stream in discrete snapshots using a
fixed number of 2000 edges to balance snapshot sizes. As we use user-item interaction datasets, with
an anomalous event indicating the removal of that user from the dataset, we generate node features
as follows: for user nodes the edge feature vector of the last interaction in the snapshot is
used, for item nodes an average over all edges associated with that item is computed.
Further, model specific preprocessing steps are used on all dataset configurations.
%
For all models, we use the recommended default parameters from the original repositories, except for
the input dimensions, which are reasonably adjusted to match the feature dimension of each configuration. All
runs are repeated over 4 random seeds and we report mean $\pm$ standard deviation.

\begin{table}[bh]
\centering
\caption{Class distribution across dataset splits (70/15/15). Percentages indicate the proportion of normal and anomaly samples within each split.}
\label{tab:datasets}
\begin{tabular}{llrrrr}
\hline
Dataset & Split & Total & Normal & Anomaly \\
\hline
Reddit
    & Train & 470,713 & 470,523 (99.96\%) & 190 (0.04\%) \\
    & Val   & 100,867 & 100,785 (99.92\%) & 82 (0.08\%) \\
    & Test  & 100,867 & 100,773 (99.91\%) & 94 (0.09\%) \\
\hline
MOOC
    & Train & 288,224 & 285,247 (98.97\%) & 2,977 (1.03\%) \\
    & Val   & 61,762  & 61,252 (99.17\%)  & 510 (0.83\%) \\
    & Test  & 61,763  & 61,184 (99.06\%)  & 579 (0.94\%) \\
\hline
Wikipedia
    & Train & 110,232 & 110,076 (99.86\%) & 156 (0.14\%) \\
    & Val   & 23,621  & 23,604 (99.93\%)  & 17 (0.07\%) \\
    & Test  & 23,621  & 23,577 (99.81\%)  & 44 (0.19\%) \\
\hline
\end{tabular}
\end{table}

To summarize, this setup allows us to answer three questions: (1) does statistical
augmentation improve detection performance? (2) which architectures benefit most
from the augmentation, and is this effect consistent across datasets? (3) does the
encoding strategy further improve performance and alter the contribution of statistical
features compared to standard augmentation?

\section{Results}\label{sec5}
We evaluate the impact on model performance (ROC-AUC) and interpretability (SHAP importance
share between embedding and statistics). Table~\ref{tab:performance} reports the detection performance of all models across
the three feature configurations on all datasets. Table~\ref{tab:shap} then breaks down the
importance share of the statistical features relative to the original embedding.

Table~\ref{tab:performance} shows a consistent and often substantial improvement from the
augmentation. Across all three datasets, at least one augmented configuration outperforms the
baseline for every model except DyGFormer on the MOOC dataset, which exhibits a small
degradation of $-0.42\%$. The most striking gains occur on the Reddit dataset, where
GraphSAGE improves by $+21.18\%$ (from $62.98\%$ to $84.15\%$ ROC-AUC) through the
$+Stats +Encoded$ configuration.
\begin{table}[t]
	\centering
	\caption{Performance comparison of anomaly detection models across feature configurations (ROC-AUC
		in \%, mean$\pm$std over 4 runs with different seeds).}
  \label{tab:performance}
	\begin{tabular}{llcccc}
		\toprule
		Dataset                          & Model                            & Original                                  & +Stats         & +Stats +Encoded                  & Max Gain                            \\
		\midrule
		\multirow{7}{*}{Reddit}          & GraphSAGE~\cite{graphsage}       & $62.98\pm2.84$                            & $61.86\pm1.92$ & \underline{$\mathbf{84.15}\pm\mathbf{1.00}$} & \cellcolor{green!50}$+21.18\pm3.37$ \\
		                                 & StrGNN~\cite{strgnn}             & $59.68\pm3.45$                            &
		$\mathbf{61.72}\pm\mathbf{2.25}$ & $60.83\pm2.79$                   &
		\cellcolor{green!17}$+3.87\pm3.85$                                                                                                                                                                        \\
		                                 & AER~\cite{aer}                   & $72.95\pm2.41$                            & $72.91\pm1.57$
		                                 & $\mathbf{79.62\pm0.77}$          & \cellcolor{green!27}$+6.67\pm2.51$                                                                                                  \\
		                                 & SAD~\cite{sad}                   & $69.26\pm1.17$                            & $69.80\pm1.99$
		                                 & $\mathbf{73.35}\pm\mathbf{0.67}$ & \cellcolor{green!16}$+4.09\pm1.41$                                                                                                  \\
		                                 & TGAT~\cite{tgat}                 & $61.48\pm0.37$                            & $61.40\pm0.38$
		                                 & $\mathbf{72.68}\pm\mathbf{0.28}$ & \cellcolor{green!40}$+11.20\pm0.44$                                                                                                 \\
		                                 & TGN~\cite{tgn}                   & $60.96\pm0.85$                            & $61.58\pm0.66$
		                                 & $\mathbf{67.88}\pm\mathbf{0.44}$ & \cellcolor{green!27}$+6.91\pm1.01$                                                                                                  \\
		                                 & DyGFormer~\cite{dygformer}       & $68.01\pm2.45$                            & $67.58\pm1.38$
		                                 & $\mathbf{74.63}\pm\mathbf{0.69}$ &
                                     \cellcolor{green!25}$+6.61\pm2.99$
                                     \\\midrule
		\multirow{7}{*}{Wikipedia}       & GraphSAGE~\cite{graphsage}                       & $72.61\pm1.40$                            & $69.33\pm1.56$
		                                 & $\mathbf{76.31}\pm\mathbf{1.14}$ & \cellcolor{green!16}$+3.70\pm0.71$                                                                                                  \\
		                                 & StrGNN~\cite{strgnn}                       & $63.56\pm2.52$                            & $66.57\pm2.69$
		                                 & $\mathbf{67.91}\pm\mathbf{2.46}$ & \cellcolor{green!26}$+6.08\pm2.82$                                                                                                  \\
		                                 & AER~\cite{aer}                             & $68.86\pm1.77$                            & $71.22\pm0.88$
		                                 & $\mathbf{76.65\pm0.62}$          & \cellcolor{green!30}$+7.79\pm1.73       $                                                                                           \\
		                                 & SAD~\cite{sad}                             & $84.77\pm1.56$                            & $84.85\pm1.01$
		                                 & \underline{$\mathbf{88.85}\pm\mathbf{0.47}$} & \cellcolor{green!17}$+4.08\pm1.72$                                                                                                  \\
                                      & TGAT~\cite{tgat} & $76.85\pm6.03$ & $\mathbf{81.87}\pm\mathbf{2.76}$ & $75.72\pm0.54$ & \cellcolor{green!17}$+5.02\pm5.13$ \\
		                                 & TGN~\cite{tgn}                              & $78.46\pm1.80$                            & $78.99\pm1.30$ & $\mathbf{82.01}\pm\mathbf{0.20}$ & \cellcolor{green!15}$+3.54\pm1.61$  \\
		                                 & DyGFormer\cite{dygformer}                        & $83.36\pm2.02$
                                     & $81.70\pm2.84$ & $\mathbf{86.92}\pm\mathbf{1.64}$ &
                                     \cellcolor{green!15}$+3.56\pm2.15$  \\\midrule
		\multirow{7}{*}{MOOC}            & GraphSAGE~\cite{graphsage}                        & $56.51\pm0.08$                            &
		$\mathbf{71.13}\pm\mathbf{0.39}$ & $69.34\pm0.48$                   & \cellcolor{green!49}$+14.62\pm0.34$                                                                                                 \\
		                                 & StrGNN~\cite{strgnn}                       & $66.48\pm0.87$                            & $66.44\pm2.60$
		                                 & $\mathbf{67.63}\pm\mathbf{0.79}$ & \cellcolor{green!11}$+1.46\pm1.29$                                                                                                  \\
		                                 & AER~\cite{aer}                             & $76.69\pm0.43$                            &
		\underline{$\mathbf{84.45\pm0.55}$}          & $84.23\pm0.28$                   &
		\cellcolor{green!30}$+7.91\pm0.56$                                                                                                                                                                        \\
		                                 & SAD~\cite{sad}                             & $67.77\pm2.37$                            & $67.76\pm7.59$
		                                 & $\mathbf{71.77}\pm\mathbf{0.34}$ & \cellcolor{green!19}$+4.44\pm2.23$                                                                                                  \\
		                                 & TGAT~\cite{tgat}                             & $60.08\pm0.34$                            & $70.91\pm0.12$
		                                 & $\mathbf{72.05}\pm\mathbf{0.09}$ & \cellcolor{green!43}$+11.98\pm0.37$                                                                                                 \\
		                                 & TGN~\cite{tgn}                              & $63.46\pm0.51$                            & $69.58\pm0.34$
		                                 & $\mathbf{71.90}\pm\mathbf{0.04}$ & \cellcolor{green!31}$+8.44\pm0.55$                                                                                                  \\
		                                 & DyGFormer~\cite{dygformer}                       &
		$\mathbf{76.71}\pm\mathbf{0.62}$ & $75.72\pm0.63$                   &
		$76.10\pm0.20$                   & \cellcolor{red!5}$-0.42\pm0.76$                                                                                                                                        \\
		\bottomrule
	\end{tabular}
\end{table}

Three patterns emerge from Table~\ref{tab:performance}. First, the augmentation consistently
improves performance on at least one dataset for every model, but the gain varies strongly across
datasets and models. This indicates that the anomaly generation process differs between domains and is
reflected in our statistics to different degrees. Second, the gain of the augmentation varies
between models and is not consistent over the discrete- nor the continuous-time category.
Third, the magnitude of the gain is not correlated with the baseline performance: both
low-performing and already high-performing models benefit from the augmentation sometimes to a
similar degree. This indicates that the statistics are not merely used to close a modelling deficiency of
weaker models but provide information that is orthogonal to existing features.
\begin{table}[t]
	\centering
	\caption{Share of SHAP importance between embeddings and statistics (mean$\pm$std over 4 runs with different seeds, in \%).}
    \label{tab:shap}
	\begin{tabular}{llcccc}
		\toprule
		\multirow{2}{*}{Dataset}   & \multirow{2}{*}{Model}   & \multicolumn{2}{c}{+Stats} &
		\multicolumn{2}{c}{+Stats +Encoded}                                                                                                                                 \\
		                           &                          & Emb                          & Stat                     & Emb                      & Stat                     \\
		\midrule
		\multirow{7}{*}{Reddit}    & GraphSAGE~\cite{graphsage}               & $70.9\pm2.5$                 & $29.1\pm2.5$
                               & \underline{$\mathbf{20.0}\pm\mathbf{0.0}$} & \underline{$\mathbf{80.0}\pm\mathbf{0.0}$}                                                                                      \\
                               & StrGNN~\cite{strgnn}                   & $\mathbf{65.2}\pm\mathbf{1.7}$     &
                               $\mathbf{34.8}\pm\mathbf{1.7}$ & $49.1\pm1.7$             & $50.9\pm1.7$             \\
		                           & AER~\cite{aer}                     & $50.0\pm1.1$                          & $50.0\pm 1.1$                       & $\mathbf{45.2\pm1.5}$                       & $\mathbf{54.8\pm1.5}$                       \\
		                           & SAD~\cite{sad}                     & $80.0\pm6.1$                 &
    $20.0\pm6.1$               & $\mathbf{34.3}\pm\mathbf{0.9}$ &
    $\mathbf{65.7}\pm\mathbf{0.9}$                                                                                                                                              \\
                               & TGAT~\cite{tgat}                    & $80.3\pm0.8$                 &
                               $19.7\pm0.8$             & $\mathbf{46.4}\pm\mathbf{2.6}$ &
                               $\mathbf{53.6}\pm\mathbf{2.6}$             \\
		                           & TGN~\cite{tgn}                     & $82.9\pm2.4$                 &
                               $17.1\pm2.4$             & $\mathbf{43.7}\pm\mathbf{2.7}$             &
                               $\mathbf{56.3}\pm\mathbf{2.7}$             \\
		                           & DyGFormer~\cite{dygformer}               & $63.6\pm8.2$                 &
                               $36.4\pm8.2$             & $\mathbf{46.2}\pm\mathbf{4.2}$             &
                               $\mathbf{53.8}\pm\mathbf{4.2}$             \\\midrule
		\multirow{7}{*}{Wikipedia} & GraphSAGE~\cite{graphsage}               & $55.0\pm5.7$                 &
    $45.0\pm5.7$               & $\mathbf{18.8}\pm\mathbf{3.4}$ & $\mathbf{81.2}\pm\mathbf{3.4}$                                                                                      \\
		                           & StrGNN~\cite{strgnn}                  & $45.5\pm3.7$                 &
                               $54.5\pm3.7$             & $\mathbf{54.2}\pm\mathbf{2.3}$ &
                               $\mathbf{45.8}\pm\mathbf{2.3}$ \\
		                           & AER~\cite{aer}                     & $65.2\pm1.7$                         & $34.8\pm1.7$                      & $\mathbf{50.8\pm3.6}$                       & $\mathbf{49.2\pm3.6}$                       \\
		                           & SAD~\cite{sad}                      & $72.9\pm3.8$                 &
    $27.1\pm3.8$               & \underline{$\mathbf{11.4}\pm\mathbf{1.3}$} &
    \underline{$\mathbf{88.6}\pm\mathbf{1.3}$}                                                                                                                                              \\
                                  & TGAT~\cite{tgat} & $\mathbf{47.6}\pm\mathbf{13.8}$ & $\mathbf{52.4}\pm\mathbf{13.8}$ & $6.0\pm0.9$ & $94.0\pm0.9$ \\
		                           & TGN~\cite{tgn}                      & $55.9\pm11.1$                &
                               $44.2\pm11.1$            & $\mathbf{9.8}\pm\mathbf{1.5}$              &
                               $\mathbf{90.2}\pm\mathbf{1.5}$             \\
		                           & DyGFormer~\cite{dygformer}               & $49.1\pm20.5$                &
                               $50.9\pm20.5$            & $\mathbf{9.9}\pm\mathbf{2.3}$              &
                               $\mathbf{90.0}\pm\mathbf{2.3}$             \\\midrule

    \multirow{7}{*}{MOOC}      & GraphSAGE~\cite{graphsage}                & $\mathbf{40.5}\pm\mathbf{6.7}$                 &
    $\mathbf{59.5}\pm\mathbf{6.7}$               & $35.2\pm6.4$ & $64.8\pm6.4$                                                                                      \\
		                           & StrGNN~\cite{strgnn}                   & $27.2\pm3.9$                 &
                               $72.8\pm3.9$             & $\mathbf{64.8}\pm\mathbf{3.8}$ &
                               $\mathbf{35.2}\pm\mathbf{3.8}$ \\
		                           & AER~\cite{aer}                      & \underline{$\mathbf{78.8\pm2.1}$}                           & \underline{$\mathbf{21.2\pm2.1}$}                       & $52.2\pm1.3$                       & $47.8\pm1.3$                       \\
		                           & SAD~\cite{sad}                     & $47.5\pm7.4$                 &
    $52.5\pm7.4$               & $\mathbf{31.4}\pm\mathbf{8.7}$ &
    $\mathbf{68.6}\pm\mathbf{8.7}$                                                                                                                                              \\
		                           & TGAT~\cite{tgat}                     & $35.3\pm9.4$                 &
                               $64.7\pm9.4$             & $\mathbf{20.4}\pm\mathbf{2.0}$            &
                               $\mathbf{79.6}\pm\mathbf{2.0}$             \\
		                           & TGN~\cite{tgn}                      & $36.8\pm17.4$                &
                               $63.2\pm17.4$            & $\mathbf{20.0}\pm\mathbf{2.1}$             &
                               $\mathbf{80.0}\pm\mathbf{2.1}$             \\
                               & DyGFormer~\cite{dygformer} & $84.9\pm4.6$ & $15.1\pm4.6$ & $43.6\pm9.8$ & $56.4\pm9.8$ \\
		\bottomrule
	\end{tabular}
\end{table}

Beyond detection performance, we analyze the extent to which the model relies on the
statistical features for its decision. Table~\ref{tab:shap} reports the SHAP importance
share of the statistical dimensions versus the original embedding features.

Table~\ref{tab:shap} reveals two main trends. First, the +Stats +Encoded configuration generally shifts
SHAP importance toward statistical features, particularly on Wikipedia, where statistics account for
more than $80\%$ of the importance for most models. On Reddit and MOOC, the effect is more
model-dependent, with several models showing a more balanced contribution.

Second, the nine statistical features often contribute substantially more than their dimensionality
would suggest. 
The best performing model-variant pairs (underscored results) rely, with $80$--$88$\% on Reddit and Wikipedia, heavily on the statistical information. On the MOOC dataset, results from the best performing models shows the opposite, with statistical information as a supportive addition instead of the main source of information.
Nevertheless, this indicates that the
statistics can provide substantial predictive information beyond the embeddings.

The encoded configuration also often reduces seed variability, especially on Wikipedia and Reddit,
but this effect is not consistent across all datasets. For example, the standard deviation decreases
for Reddit DyGFormer from $8.2\%$ to $4.2\%$, but increases for MOOC DyGFormer from
$4.6\%$ to $9.8\%$. Thus, encoding changes the balance of feature importance and can improve stability,
but neither effect is universal.

To summarize the results with respect to the experimental questions from Section~\ref{sec4} we observe the following:
\begin{enumerate}
\item In general, statistical augmentation improves detection performance in the majority of the applied configurations, often by
substantial margins.
\item Overall, no architecture consistently benefits the most across datasets, indicating
that the effect of augmentation is dataset- and model-dependent.
\item Furthermore, the encoded configuration
often tends to outperform the concatenation strategy, shifts SHAP importance toward the statistical features and improves stability, although these effects are not consistent across all datasets and models.
\end{enumerate}

\section{Discussion}\label{sec6}
Our results indicate that statistical feature augmentation provides information complementary to existing dynamic graph representations. Performance improves for most models, but the magnitude of
the gain varies across datasets and architectures. This suggests that the applicability and usefulness of short-term behavioral statistics depends on the characteristics of the underlying interaction data rather than on a specific model family.

In general, the results of the SHAP analysis supports this interpretation. The nine statistical features often account for a substantial share of decision importance despite their small dimensionality. The encoded configuration generally increases their contribution, particularly on Wikipedia, while Reddit and MOOC show more model-dependent effects. Thus, the augmentation not only improves performance in many settings but also makes the contribution of specific behavioral statistics more directly
interpretable.

\section{Conclusion}\label{sec7}
In this paper we showed the positive impact of enriching dynamic network data with interaction
statistics. We argue, that augmenting the data with classical statistics enables us to
learn network dynamics in a simplified manner and on top, it can be used as a strong basis for the
explainability aspect of these models.
Therefore, with this work we want to highlight the broader question: What kind of interaction behavior can be moved from the learning task in the input space to increase performance and interpretability.

For future work, we aim to address this question by increasing the set of statistics, especially focusing on more complex interaction patterns that are grounded in the theory of social analysis. Our work showed the impact of an additional projection into a combined feature space of embeddings and statistics, which should also be researched further.
In addition, we aim to extend our approach for more datasets. Here, also an analysis of causal relationships becomes important, since large standard deviations of SHAP importance may indicate an association with the training process. Future work should therefore examine specifically additional more complex behavioral statistics and a broader range of datasets.

\bibliographystyle{spmpsci} 
\bibliography{refs} 

@article{sad,
  title = {Sad: Semi-supervised anomaly detection on dynamic graphs},
  author = {Tian, Sheng and Dong, Jihai and Li, Jintang and Zhao, Wenlong and Xu
            , Xiaolong and Song, Bowen and Meng, Changhua and Zhang, Tianyi and
            Chen, Liang and others},
  journal = {arXiv preprint arXiv:2305.13573},
  year = {2023},
}

@inproceedings{graphsage,
  author = {Hamilton, William L. and Ying, Rex and Leskovec, Jure},
  title = {Inductive Representation Learning on Large Graphs},
  booktitle = {Proc. NIPS},
  year = {2017},
}

@inproceedings{strgnn,
  title = {Structural temporal graph neural networks for anomaly detection in
           dynamic graphs},
  author = {Cai, Lei and Chen, Zhengzhang and Luo, Chen and Gui, Jiaping and Ni,
            Jingchao and Li, Ding and Chen, Haifeng},
  booktitle = {Proc. ACM CIKM},
  pages = {3747--3756},
  year = {2021},
}

@article{aer,
  title = {Anonymous edge representation for inductive anomaly detection in
           dynamic bipartite graph},
  author = {Fang, Lanting and Feng, Kaiyu and Gui, Jie and Feng, Shanshan and Hu
            , Aiqun},
  journal = {Proc. VLDB Endowment},
  volume = {16},
  number = {5},
  pages = {1154--1167},
  year = {2023},
  publisher = {VLDB Endowment},
}

@article{tgat,
  title = {Inductive representation learning on temporal graphs},
  author = {Xu, Da and Ruan, Chuanwei and Korpeoglu, Evren and Kumar, Sushant
            and Achan, Kannan},
  journal = {arXiv preprint arXiv:2002.07962},
  year = {2020},
}

@article{tgn,
  title = {Temporal graph networks for deep learning on dynamic graphs},
  author = {Rossi, Emanuele and Chamberlain, Ben and Frasca, Fabrizio and Eynard
            , Davide and Monti, Federico and Bronstein, Michael},
  journal = {arXiv preprint arXiv:2006.10637},
  year = {2020},
}

@article{dygformer,
  title = {Towards better dynamic graph learning: New architecture and unified
           library},
  author = {Yu, Le and Sun, Leilei and Du, Bowen and Lv, Weifeng},
  journal = {Proc. NIPS},
  volume = {36},
  pages = {67686--67700},
  year = {2023},
}

@article{shap,
  title = {A unified approach to interpreting model predictions},
  author = {Lundberg, Scott M and Lee, Su-In},
  journal = {Proc. NIPS},
  volume = {30},
  year = {2017},
}

@article{rem,
  title = {4. A Relational Event Framework for Social Action},
  author = {Carter T. Butts},
  journal = {Sociol. Methodol.},
  year = {2008},
  volume = {38},
  pages = {155 - 200},
  OPTurl = {https://api.semanticscholar.org/CorpusID:120970495},
}

@inproceedings{sitgnn,
  title = {Inference of Sequential Patterns for Neural Message Passing in
           Temporal Graphs},
  author = {Jan von Pichowski and Vincenzo Perri and Lisi Qarkaxhija and Ingo
            Scholtes},
  booktitle = {Proc. Learning on Graphs Conference},
  year = {2024},
  OPTurl = {https://openreview.net/forum?id=5pAfYgmbth},
}

@article{caw,
  title = {Inductive Representation Learning in Temporal Networks via Causal
           Anonymous Walks},
  author = {Yanbang Wang and Yen-Yu Chang and Yunyu Liu and Jure Leskovec and
            Pan Li},
  journal = {ArXiv},
  year = {2021},
  volume = {abs/2101.05974},
  OPTurl = {https://api.semanticscholar.org/CorpusID:231627759},
}

@inproceedings{neurtws,
  title = {Neural Temporal Walks: Motif-Aware Representation Learning on
           Continuous-Time Dynamic Graphs},
  author = {Ming Jin and Yuan-Fang Li and Shirui Pan},
  booktitle = {Proc. NIPS},
  editor = {Alice H. Oh and Alekh Agarwal and Danielle Belgrave and Kyunghyun
            Cho},
  year = {2022},
  OPTurl = {https://openreview.net/forum?id=NqbktPUkZf7},
}

@article{tempme,
  title = {TempME: Towards the Explainability of Temporal Graph Neural Networks
           via Motif Discovery},
  author = {Jialin Chen and Rex Ying},
  journal = {ArXiv},
  year = {2023},
  volume = {abs/2310.19324},
  OPTurl = {https://api.semanticscholar.org/CorpusID:264829176},
}

@article{Alomrani2022DyG2VecRL,
  title = {DyG2Vec: Representation Learning for Dynamic Graphs with
           Self-Supervision},
  author = {Mohammad Ali Alomrani and Mahdi Biparva and Yingxue Zhang and Mark
            Coates},
  journal = {ArXiv},
  year = {2022},
  volume = {abs/2210.16906},
  OPTurl = {https://api.semanticscholar.org/CorpusID:261959483},
}

@article{stream,
    author = {Filippi-Mazzola, Edoardo and Wit, Ernst C},
    title = {A stochastic gradient relational event additive model for modelling US patent citations from 1976 to 2022},
    journal = {J. R. Stat. Soc. C Appl. Stat.},
    volume = {73},
    number = {4},
    pages = {1008-1024},
    year = {2024},
    month = {08},
    issn = {0035-9254},
    doi = {10.1093/jrsssc/qlae023},
    OPTurl = {https://doi.org/10.1093/jrsssc/qlae023},
    eprint = {https://academic.oup.com/jrsssc/article-pdf/73/4/1008/58805408/qlae023.pdf},
}

@article{dream,
  title={Modeling non-linear Effects with Neural Networks in Relational Event Models},
  author={Edoardo Filippi-Mazzola and Ernst C Wit},
  journal={Soc. Networks},
  year={2023},
  volume={79},
  pages={25-33},
  url={https://api.semanticscholar.org/CorpusID:266362538}
}

@inproceedings{Xia2023ExplainingTG,
  title={Explaining Temporal Graph Models through an Explorer-Navigator Framework},
  author={Wenwen Xia and Mincai Lai and Caihua Shan and Yaofang Zhang and Xinnan Dai and Xiang Li and Dongsheng Li},
  booktitle={Proc. ICLR},
  year={2023},
  url={https://api.semanticscholar.org/CorpusID:259298210}
}

@article{interdonato2019feature,
  title={Feature-rich networks: going beyond complex network topologies},
  author={Interdonato, Roberto and Atzmueller, Martin and Gaito, Sabrina and Kanawati, Rushed and Largeron, Christine and Sala, Alessandra},
  journal={Appl. Netw. Sci.},
  volume={4},
  number={1},
  pages={1--13},
  year={2019},
  publisher={Springer}
}

@Article{KC:97,
  author = 	 {Karin Knorr-Cetina},
  title = 	 {{Sociality with Objects: Social Relations in Postsocial Knowledge Societies}},
  journal = 	 {Theory Cult. Soc.},
  year = 	 {1997},
  OPTkey = 	 {},
  volume = 	 {14},
  number = 	 {4},
  pages = 	 {1--43},
  OPTmonth = 	 {},
  OPTnote = 	 {},
  OPTannote = 	 {}
}

@article{Atzmueller:14:CoRR,
  author    = {Martin Atzmueller},
  title     = {{Data Mining on Social Interaction Networks}},
  journal   = {JDMDH},
  volume    = {1},
  month     = {June},
  year      = {2014},
  OPTee        = {http://arxiv.org/abs/1312.6675},
  OPTbibsource = {DBLP, http://dblp.uni-trier.de}
}

@article{holme2012temporal,
  title={Temporal networks},
  author={Holme, Petter and Saram{\"a}ki, Jari},
  journal={Phys. Rep.},
  volume={519},
  number={3},
  pages={97--125},
  year={2012},
  publisher={Elsevier}
}

@book{WF:94,
  author = {Wasserman, Stanley and Faust, Katherine},
  edition = 1,
  isbn = {9780521387071},
  OPTnumber = 8,
  publisher = {Cambridge University Press},
  series = {Structural Analysis in the Social Sciences},
  title = {{Social Network Analysis: Methods and Applications}},
  year = 1994
}

@article{goldenberg2010survey,
  title={A survey of statistical network models},
  author={Goldenberg, Anna and Zheng, Alice X and Fienberg, Stephen E and Airoldi, Edoardo M},
  journal={Found. Trends Mach. Learn.},
  volume={2},
  number={2},
  pages={129--233},
  year={2010},
  publisher={Emerald Publishing Limited}
}

@article{snijders2001statistical,
  title={The statistical evaluation of social network dynamics},
  author={Snijders, Tom AB},
  journal={Sociol. Methodol.},
  volume={31},
  number={1},
  pages={361--395},
  year={2001},
  publisher={Wiley Online Library}
}

@article{carley2007toward,
  title={Toward an interoperable dynamic network analysis toolkit},
  author={Carley, Kathleen M and Diesner, Jana and Reminga, Jeffrey and Tsvetovat, Maksim},
  journal={Decis. Support Syst.},
  volume={43},
  number={4},
  pages={1324--1347},
  year={2007},
  publisher={Elsevier}
}

@article{Savage:14,
  title={{Anomaly Detection in Online Social Networks}},
  author={Savage, David and Zhang, Xiuzhen and Yu, Xinghuo and Chou, Pauline and Wang, Qingmai},
  journal={Soc. Networks},
  volume={39},
  pages={62--70},
  year={2014},
  publisher={Elsevier}
}

@Article{Akoglu:15,
  author = 	 {Akoglu, L and Tong, H and Koutra, D},
  title = 	 {{Graph Based Anomaly Detection and Description}},
  journal = 	 {Data. Min. Knowl. Disc.},
  year = 	 {2015}, 
 volume = {29},
 number = {3},
 month = may,
 year = {2015},
 OPTissn = {1384-5810},
 pages = {626--688},
 numpages = {63},
 OPTurl = {http://dx.doi.org/10.1007/s10618-014-0365-y},
 OPTdoi = {10.1007/s10618-014-0365-y},
 publisher = {Kluwer Academic Publishers},
 address = {Hingham, MA, USA}, 
  OPTmonth = 	 {},
  OPTnote = 	 {},
  OPTannote = 	 {}
}

@article{roth2008explanation,
  title={On Explanation.},
  author={Roth-Berghofer, Thomas and Richter, Michael M},
  journal={K{\"u}nstliche Intell.},
  volume={22},
  number={2},
  pages={5--7},
  year={2008}
}

@article{gunning2017explainable,
  title = {{XAI - Explainable Artificial Intelligence}},
  author = {David Gunning  and Mark Stefik  and Jaesik Choi  and Timothy Miller  and Simone Stumpf  and Guang-Zhong Yang},
  journal = {Sci. Robot.},
  volume = {4},
  OPTnumber = {37},
  year = {2019},
  OPTpages = {eaay7120},
}

@article{AFKS:24,
  title={Explainable and interpretable machine learning and data mining},
  author={Atzmueller, Martin and F{\"u}rnkranz, Johannes and Kliegr, Tom{\'a}{\v{s}} and Schmid, Ute},
  journal={Data Min Knowl Disc},
  volume={38},
  number={5},
  pages={2571--2595},
  year={2024},
  publisher={Springer},
  OPTdoi = {10.1007/s10618-024-01041-y},
}

@article{barredo2019explainable,
  author = {Alejandro {Barredo Arrieta} and Natalia Díaz-Rodríguez and Javier {Del Ser} and Adrien Bennetot and Siham Tabik and Alberto Barbado and Salvador Garcia and Sergio Gil-Lopez and Daniel Molina and Richard Benjamins and Raja Chatila and Francisco Herrera},
  title = {{Explainable Artificial Intelligence (XAI)}: Concepts, Taxonomies, Opportunities and Challenges Toward Responsible {AI}},
  journal = {Inf. Fusion},
  volume = {58},
  pages = {82--115},
  year = {2020},
}
\end{document}